# Heterojunction interface regulation to realize high-performance flexible Kesterite solar cells

Xiao Xu†, Jiazheng Zhou†, Kang Yin, Jinlin Wang, Licheng Lou, Dongmei Li, Jiangjian Shi*, Huijue Wu, Yanhong Luo*, and Qingbo Meng*

* Corresponding authors.
† These authors contribute equally.

X. Xu, J. Zhou, K. Yin, J. Wang, L. Lou, D. Li, J. Shi, H. Wu, Y. Luo, Q. Meng
Beijing National Laboratory for Condensed Matter Physics, Institute of Physics, Chinese Academy of Sciences (CAS), Beijing 100190, P. R. China
E-mail: qbmeng@iphy.ac.cn; yhluo@iphy.ac.cn; shijj@iphy.ac.cn

Q. Meng
Center of Materials Science and Optoelectronics Engineering, University of Chinese Academy of Sciences, Beijing 100049, P. R. China

X. Xu, J. Zhou, K. Yin, J. Wang, L. Lou, D. Li, Y. Luo
School of Physical Sciences, University of Chinese Academy of Sciences, Beijing 100049, P. R. China

D. Li, Y. Luo, Q. Meng
Songshan Lake Materials Laboratory, Dongguan, Guangdong 523808, P. R. China

**Abstract**

Flexible $Cu_2ZnSn(S, Se)_4$ (CZTSSe) solar cells take the advantages of environmental friendliness, low cost, and multi-scenario applications, and have drawn extensive attention in recent years. Compared with rigid devices, the lack of alkali metal elements in the flexible substrate is the main factor limiting the performance of flexible CZTSSe solar cells. This work proposes a Rb ion additive strategy to simultaneously regulate the CZTSSe film surface properties and the CdS chemical bath deposition (CBD) processes. Material and chemical characterization reveals that Rb ions can passivate the detrimental $Se^0$ cluster defect and additionally provide a more active surface for the CdS epitaxial growth. Furthermore, Rb can also coordinate with thiourea (TU) in the CBD solution and improve the ion-by-ion deposition of the CdS layer. Finally, the flexible CZTSSe cell fabricated by this strategy has reached a high total-area efficiency of 12.63% (active-area efficiency of 13.2%), with its $V_{OC}$ and FF reaching 538 mV and 0.70, respectively. This work enriches the alkali metal passivation strategies and provides new ideas for further improving flexible CZTSSe solar cells in the future.

## Introduction

Flexible thin-film solar cells have important application prospects in the areas of building integrated, wearable, and portable photovoltaics, benefiting from the lightweight and bendable advantages.[1] Flexible solar cells have been realized in a series of photovoltaic materials, including the organic-inorganic hybrid perovskite, organic molecules or polymers, and inorganic GaAs, copper indium gallium selenium (CIGS) and CdTe.[2-6] However, the flexible solar cells still have not yet been marketed in a large scale, mainly due to issues including the high manufacturing cost, toxic composition and relatively low stability. Kesterite $Cu_2ZnSn(S, Se)_4$ (CZTSSe) composed of earth-abundant elements is an emerging inorganic material developed as a promising alternative to traditional CIGS and CdTe, having advantages of low cost, high material stability, tunable bandgap, and high industrial compatibility.[7,8] These advantages of CZTSSe will help overcome current problems of flexible solar cells, and thus promote their larger-scale commercialization.

In the past ten years, the research on the flexible CZTSSe solar cell has achieved considerable progress, with photoelectric conversion efficiency being improved from 2.4% to about 11.2%.[9-16] Nevertheless, this efficiency still has a large gap to that of current other flexible solar cells and is also obviously lower than the rigid CZTSSe solar cell fabricated on a SLG/Mo (SLG: soda lime glass) substrate. The lack of alkali metal elements (e.g. Na) in the flexible substrate such as the Mo or stainless-steel foil could be the main origin causing the performance difference between the flexible and the rigid CZTSSe solar cells. This is because that alkali metal elements are usually believed to be an important factor promoting the CZTSSe crystallization, introducing carrier doping, and passivating defects.[17-22] To compensate the alkali metal elements in flexible devices, Li, Na and K elements have been artificially introduced into the CZTSSe absorber by pre-deposition, post-treatment or designing Na containing Mo substrate.[23-25] These strategies have effectively improved the crystal morphology of the CZTSSe film and thus resulted in device efficiency enhancement. However, compared to the steady supply from the SLG substrate, the artificial incorporation is often difficult to precisely control the content and distribution of alkali metal elements, especially there is a significant element exchange between the CZTSSe film and the surroundings during the selenization process.[26-28] Moreover the volatilization loss usually leads to alkali metal element deficiency in the surface region of the CZTSSe film. This may make the alkali metal passivation of the CZTSSe surface defects not fully realized. Therefore, although it is challenging, there is still considerable potential for utilizing alkali metal strategies to improve the cell performance.

In this work, we have explored a Rb ion additive strategy to simultaneously regulate the CZTSSe film surface properties and the CdS chemical bath deposition (CBD) processes, and thus have realized a high total-area efficiency of 12.63% (active-area efficiency of 13.2%) and excellent bending performance in flexible CZTSSe solar cells. It is found that Rb ions can break up the Se-Se bond on the CZTSSe film surface by Rb-Se interactions, which could passivate the detrimental $Se^0$ cluster defect and additionally provide a more active surface for the CdS epitaxial growth. In addition, Rb is found to be able to coordinate with the thiourea (TU) in the solution and thus improve the ion-by-ion deposition of the CdS layer, making the CdS layer much more uniform and compact. These benefits have effectively reduced the defect response, improved charge dynamics properties of the cell, and thus helped us to achieve an impressive progress in highly efficient flexible CZTSSe solar cells.

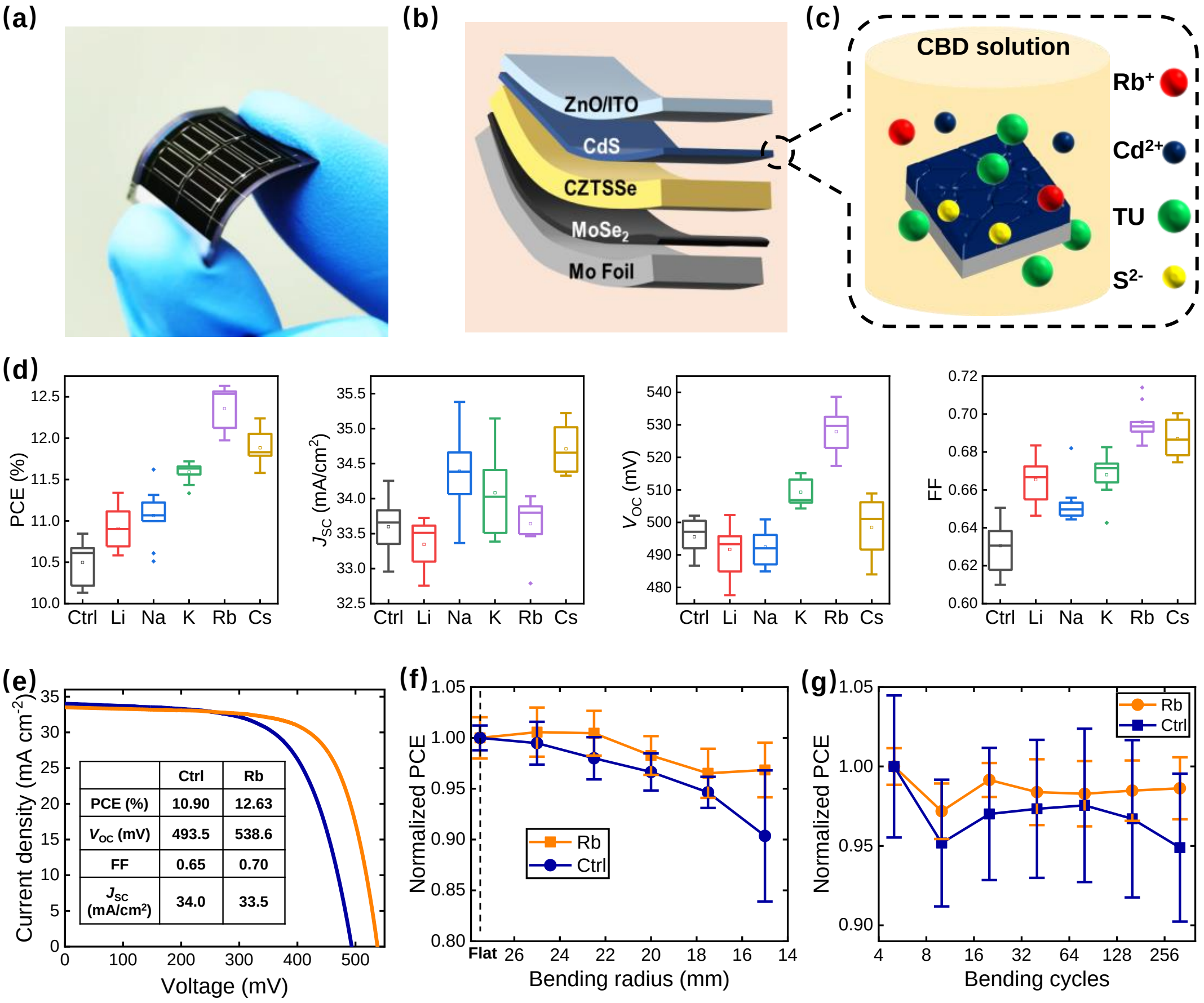

Figure 1 (a) Photo and (b) device configuration of the flexible CZTSSe solar cell. (c) Schematic diagram of the CdS CBD with adding alkali metal ions, such as $Rb^+$. (d) Statistic performance parameters of the cells modified by different alkali metal ions. (e) Current-voltage characteristics of the champion cells. (f-g) Performance evolution of the cells when bended at different radius and different bending cycles.

## Results and discussions

Figure 1(a) gives a photo of our fabricated flexible CZTSSe solar cell, which is fabricated on a Mo foil and have a similar structure to that of the rigid device, as shown in Figure 1(b). Besides the Li doping in the bulk CZTSSe through precursor solution (the impact of Li doping on device performance is shown in Figure S1), alkali metal ions are also added in the CBD solution for the CZTSSe surface regulation, as schematically shown in Figure 1(c). We have systematically optimized the alkali metal type (including Li, Na, K, Rb and Cs) and concentration, and their influence on the cell performance is shown in Figure 1(d). It can be found that all the alkali metal ions can improve the photoelectric conversion efficiency (PCE) of the cell and the highest performance is obtained when the Rb is used (labelled as Rb-cell for clarity). The average PCE of the Rb modified cell reaches 12.3%, much superior to that of the control cell (10.5%). This PCE enhancement mainly benefits from the open-circuit voltage ($V_{OC}$) and fill factor (FF) improvements. The Rb-cell has achieved a highest total-area PCE of 12.63% (active-area efficiency of 13.2%) with short-circuit current density ($J_{SC}$) of 33.5 mA cm$^{-2}$, $V_{OC}$ of 538.6 mV and FF of 0.70. Comparatively, the control cell shows much lower $V_{OC}$ (493.5 mV) and FF (0.65). More detailed parameters comparison with previously reported cells is given in Table 1. Compared to the flexible cell reported by DGIST, our Rb-cell shows similar $V_{OC}$ but a much higher FF.[9] The FF of our flexible cell is even slightly higher than the recorded result that we have obtained in the SLG based cell.[29] The external quantum efficiency spectra are shown in Figure S2.

Table 1. Cell performance parameters of our and previously reported cells.

| Device | | Substrate | PCE (%) | $V_{OC}$ (mV) | $J_{SC}$ (mA cm$^{-2}$) | FF (%) | $E_g$ (eV) | $E_g/q-V_{OC}$ (mV) |
|---|---|---|---|---|---|---|---|---|
| This work | Control | Mo foil | 10.9 | 493.5 | 34.0 | 64.7 | 1.09 | 596.5 |
| | Rb-cell | Mo foil | 12.6 | 538.6 | 33.5 | 70.0 | 1.09 | 551.4 |
| [9] | DGIST cell | Mo foil | 11.2 | 539.0 | 33.1 | 62.8 | 1.10 | 561.0 |
| [29] | IOP cell | SLG | 13.6 | 537.5 | 36.2 | 69.8 | 1.09 | 552.5 |

We also made bending tests of our flexible cells, and the results are shown in Figure 1 (f-g) and Figure S3-4. For the control cell in one-cycle bending, its PCE reduces continuously when the bending radius is decreased from infinity (pristine cell) to 15 mm. At 15 mm, only 90% of its initial average PCE is sustained. For the Rb-cell, when bending at 24 and 22 mm, the cell average PCE even shows a little increase. At smaller bending radius, the PCE also shows a continuous decrease but can still sustain 97% of the initial value, which is much superior to that of the control cell. At 22 mm, bending cycle test was further made. After 320 cycles' bending, the Rb-cell sustains 98.6% of its initial average PCE, which is also superior to the control cell. These results mean that besides the better photoelectric conversion performance, the Rb regulation has also improved the stability of the device microscopic structure, especially the CZTSSe/CdS/ZnO heterointerface bonding robustness.[23,24]

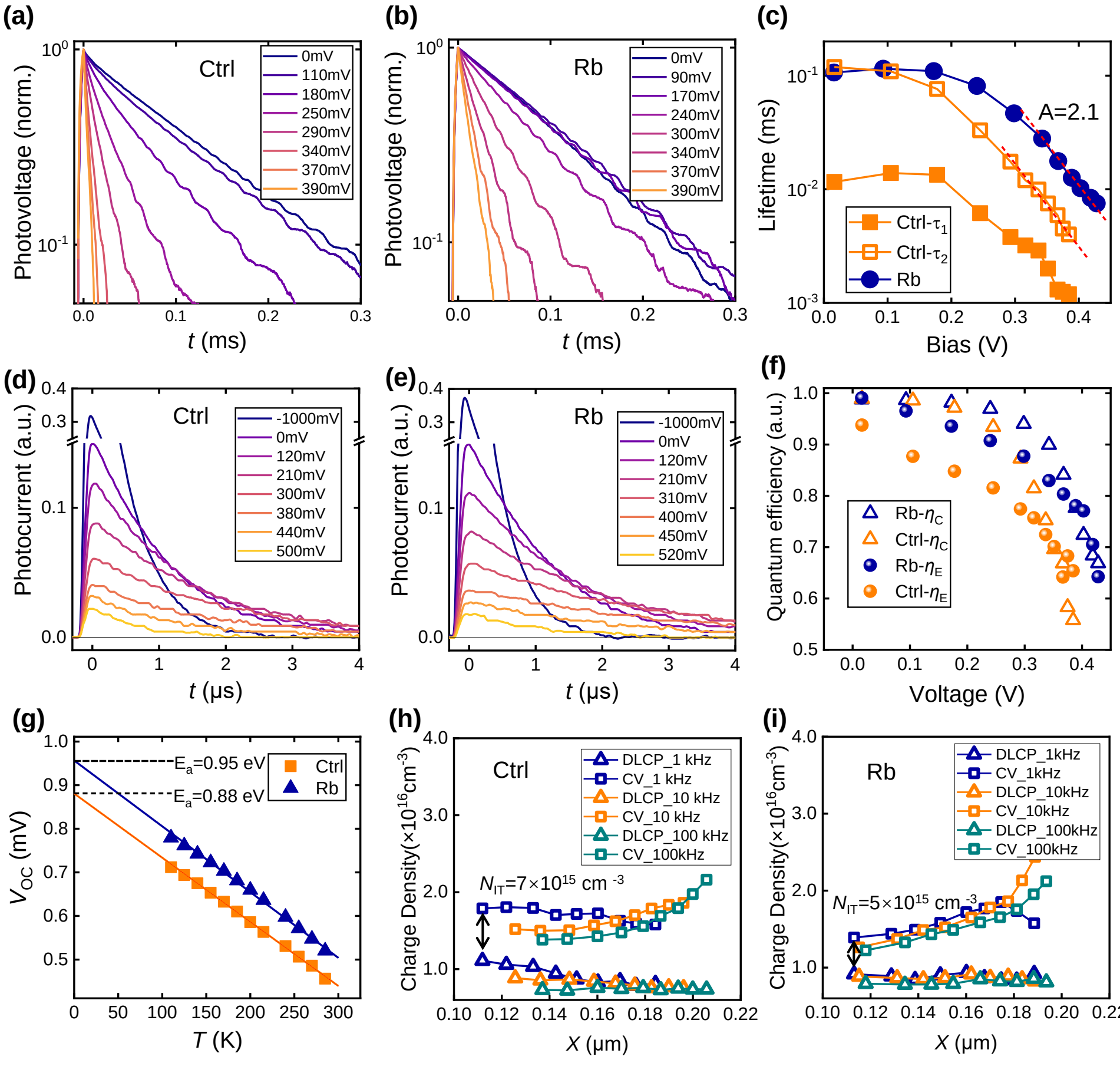

Figure 2. (a-c) Modulated transient photovoltage characteristics and the derived charge recombination lifetime of the cells. (d-f) Modulated transient photocurrent characteristics and the derived charge extraction and collection efficiency of the cells. (g) Relationship between the cell $V_{OC}$ and the temperature. (h-i) C-V and DLCP charge profiles of the cells measured

under different AC frequencies.

Charge dynamics and defect properties of the cells are firstly investigated to reveal the device physics mechanism of cell performance enhancement arisen from the Rb regulation. Through the modulated transient photovoltage, we found that the photovoltage dynamics of the Rb-cell is a little different from that of the control cell. Specifically, the control cell exhibits a double exponential photovoltage decay behavior with a fast decay in the early tens of microsecond (Figure 2(a)), while the photovoltage decay of the Rb-cell follows a single exponential character (Figure 2(b)). This means that besides the generally-existed backward charge transfer arisen from charge diffusion or recombination in the bulk absorber, severe interface charge recombination also exists in the control cell. Comparatively, the Rb regulation has effectively eliminated interface charge recombination in the cell. For clarity, the photovoltage decay lifetime of these two cells under different bias voltages have been extracted and shown in Figure 2(c). The fitting of the decay lifetime corresponding to the backward charge transfer of these two cells yields a similar ideality factor ($A$) of about 2.1.[30] This result indicates that these two cells have identical backward charge transfer mechanism, that is, bulk charge recombination in the depletion region. As such, the main difference between these two cells lies in the front interface, which agrees with the speculation obtained from the bending tests. The modulated transient photocurrent is further used to reflect the charge transport properties of the cells. As in Figure 2(d), the Rb-cell shows a faster photocurrent decay dynamics under -1 V, indicating better charge transport ability of the CZTSSe/CdS/ZnO interface region. When under high voltage of about 0.5 V, the situation has been reversed and the Rb-cell shows a slower photocurrent decay dynamics compared to that of the control cell. This arises because the charge recombination has also been involved in the photocurrent decay process under this high bias voltage. As such, the slower photocurrent decay of the Rb-cell also indicates that this cell has a reduced interface and bulk charge recombination rate, which agrees well with the photovoltage results. Based on these modulated electrical transient measurements, the charge extraction ($\eta_E$) and collection ($\eta_C$) efficiencies corresponding to the charge loss occurred in the bulk and interface transport processes, respectively, are derived and shown in Figure 2(f).[31,32] Due to the interface charge recombination, $\eta_C$ of the control cell decreases much faster than that of the Rb-cell, which finally resulted in obviously lower $V_{OC}$.

Temperature-dependent $V_{OC}$ has also been measured to reflect the interface charge loss properties.[33] As in Figure 2(g), the activation energy ($E_a$) of the Rb-cell is evaluated to be 0.95 eV, which is 70 meV higher than that of the control cell. This difference also supports that the Rb-cell has obviously smaller interface charge recombination. The defect distribution in the cell is further measured using deep level drive profile (DLCP) and capacitance-voltage (C-V) methods and specific derived data are listed in Table S1-2.[34] As in Figure 2(h-i), these two cells show similar defect charge distribution in the bulk CZTSSe while the control cell gives a more obvious frequency response in the interface region. Specifically, when decreasing the frequency from 100 to 1 kHz, the measured defect charge concentration in the interface region is obviously increased and the depletion width is also decreased. Comparatively, the defect charge distribution in the Rb-cell is little influenced by the measurement frequency. The interface defect concentration ($N_{IT}$) is further quantified using the difference between the C-V and DLCP results, which gives $7\times10^{15}$ and $5\times10^{15}$ $cm^{-3}$ of these two cells, respectively. Based on the above photoelectric characterization, it can be confirmed that the performance enhancement of the Rb-cell mainly comes from the improvement in the heterojunction interface quality.

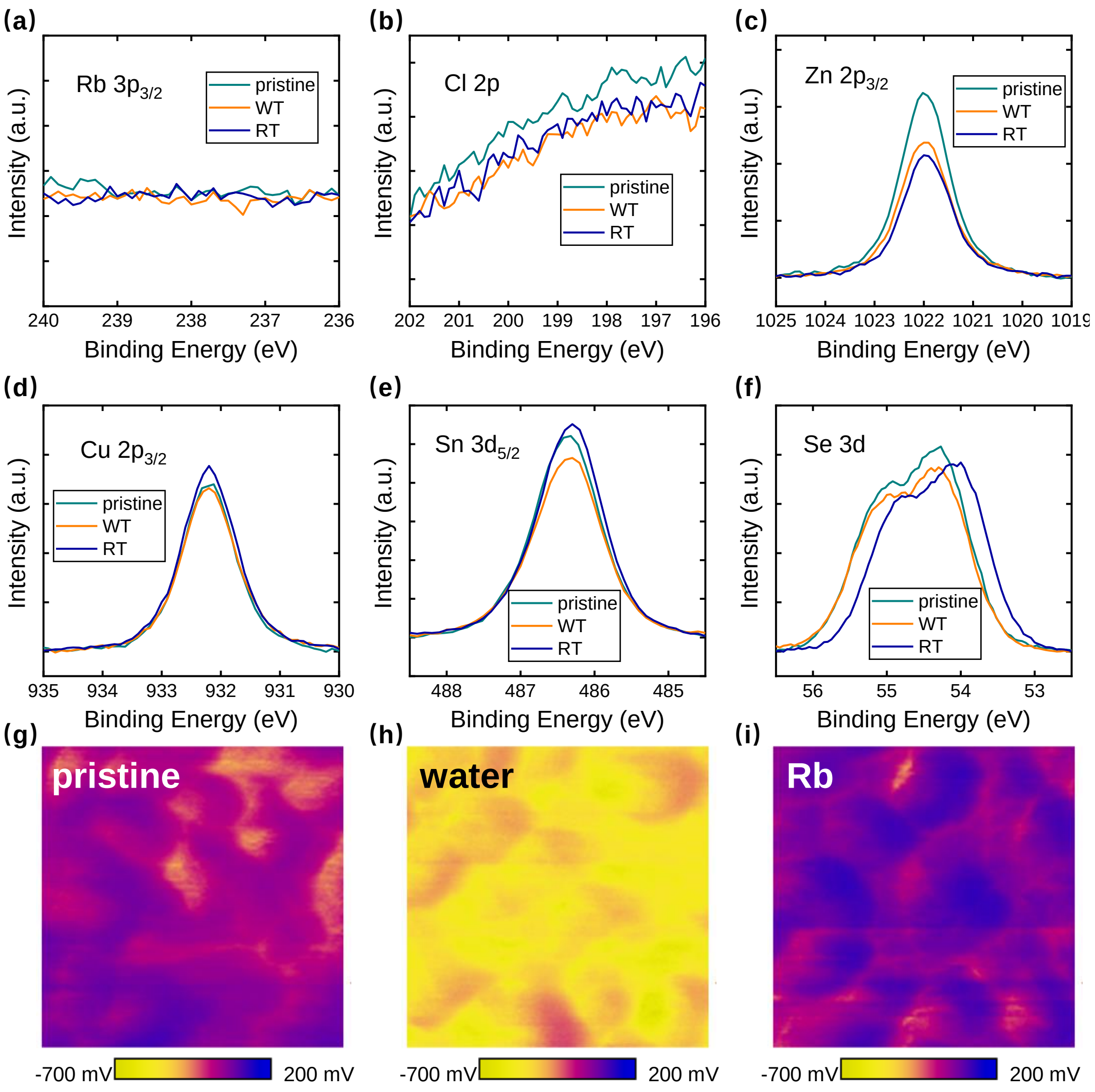

Figure 3. (a-f) XPS spectra (Rb 3p, Cl 2p, Zn 2p, Cu 2p, Sn 3d, and Se 3d) and (g-i) KPFM images of the CZTSSe films before and after treated by water or RbCl aqueous solution.

We subsequently investigate the material and chemistry mechanism of the Rb regulation on the heterojunction interface. The influence of the added RbCl on the surface of the CZTSSe film has been analyzed using X-ray photoelectron spectroscopy (XPS), as shown in Figure 2(a-f). In this study, the CZTSSe film was treated by water (80 °C, labelled as WT) or RbCl aqueous solution (labelled as RT). It is notable that no Rb can be detected on the film surface, implying that the Rb does not insert into the CZTSSe lattice or deposit on the film surface.[35] Nonetheless, it is found that XPS intensity of other elements in the film has been obviously modified after the treatment. Specifically, XPS intensity of Zn, Sn and Se elements is decreased after the water treatment.[36] This agrees with previous reports that the surface solids can be partly dissolved in the water.[37] Although this could help remove secondary phases in the film surface region to some extent, excessive dissolution will also damage the surface structure of CZTSSe, especially introducing Sn vacancy deep defect. Interestingly, when the film is treated by RbCl solution, the loss of Sn has been effectively suppressed while more Zn loss is observed. The Zn loss will help remove the ZnSe secondary phase and induce more Zn vacancies in the film surface region. In the CdS deposition process, these Zn vacancies will give more places for the $Cd^{2+}$ adsorption and diffusion, which may assist the heteroepitaxial growth.[38] More interestingly, it is found that the Rb treatment has shifted the Se 3d XPS peak to a lower energy position by about 0.3 eV, indicating more obvious anionization of Se element.[39] That is, in the pristine and WT films, a certain amount of elemental Se ($Se^0$) exists in the CZTSSe lattice terminal, forming Se-Se bond through surface reconstruction.[22,40] It has been pointed out that this surface atomic structure could introduce deep defect in the film surface. As such, the Rb treatment has helped eliminated these detrimental $Se^0$ structures. This agrees well with previous theoretical calculation result that the alkali metal ions can break up the Se-Se bond through alkali metal-Se interactions.[22] The influence on the surface electric properties is further characterized by Kelvin probe force microscope (KPFM). The contacting potential difference (CPD) mapping of the pristine and treated films is given in Figure 3(g-i). It can be seen that the water treatment has obviously decreased the surface CPD of the film, indicating upward bending of the energy band in the surface region.[41] This could arise from the Sn vacancies which usually acts as deep acceptor defect.[42-44] Comparatively, such phenomenon has not been observed in the Rb treated film because of the suppressed Sn loss. The RT film shows a little increase in the CPD compared to the pristine sample, which may arise from the change in the surface Se atomic structure. Both treatments do not influence the bulk properties according to the steady-state photoluminescence (PL) results (Figure S5).

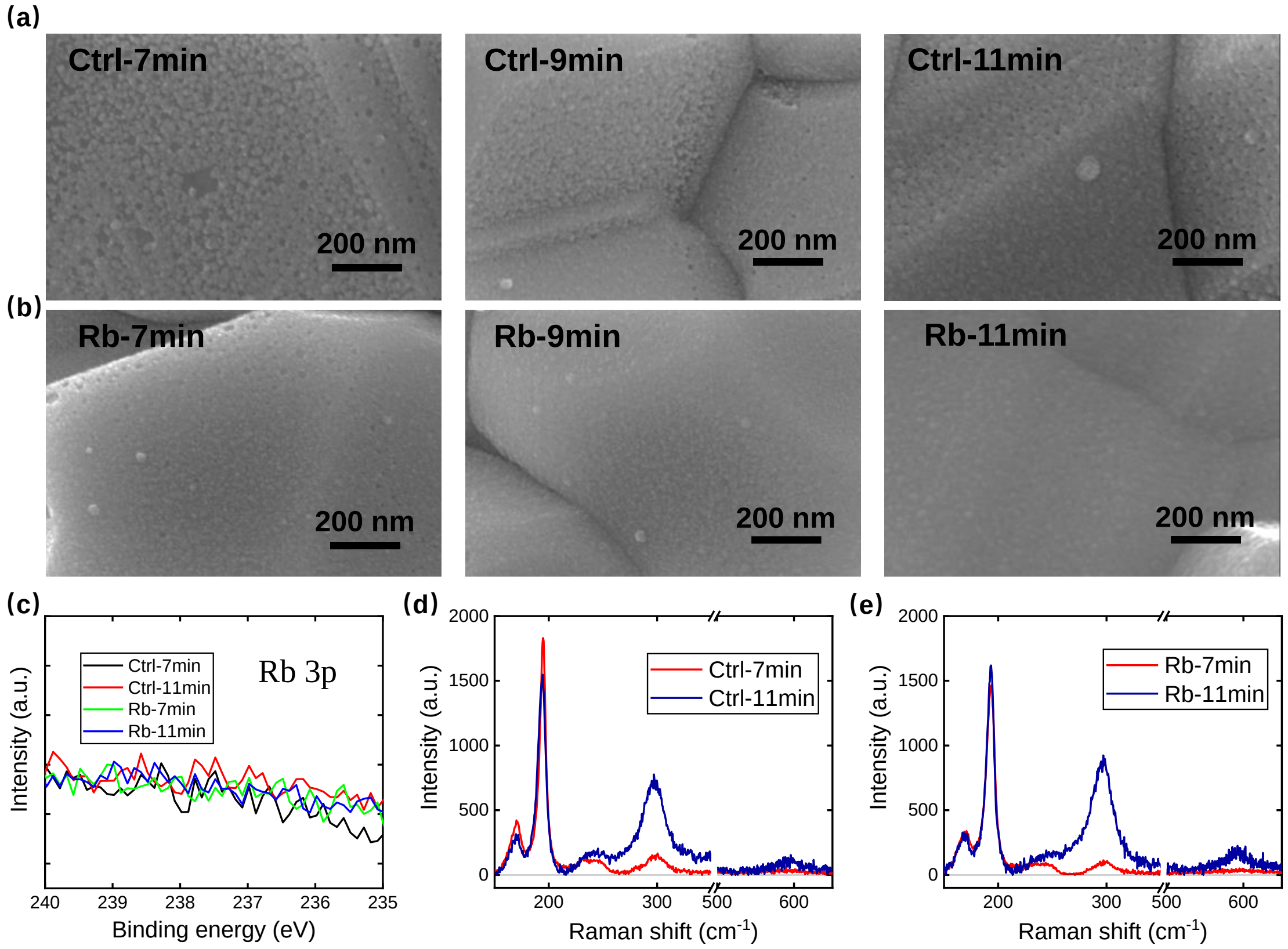

Figure 4. (a-b) Top-view SEM morphology evolution of the CdS films on CZTSSe films at different deposition time. (c) XPS spectra of the deposited CdS films in the energy range of the Rb 3p orbit. (d-e) Raman spectra of the CZTSSe/CdS films.

The influence of the Rb induced CZTSSe film surface modification on the CdS deposition process is further studied. As the top-view scanning electron microscope (SEM) images show in the Figure 4(a-b), the adding of Rb in the CBD solution has obviously changed the morphology of the deposited CdS. In the control sample, the CdS film is composed of small nanoparticles and large pinholes can be clearly seen in the film. Excess accumulation of CdS solids can also been seen in the grain boundary region. Even in the final sample (Ctrl-11min), the CdS film still exhibits rough surface. This implies that the CdS deposition in this sample does not have a good heteroepitaxial growth but mainly adopts a cluster-by-cluster deposition mode (reaction function is shown in Figure S6).[45,46] This arises mainly because Cd ions have low adsorption activity on the CZTSSe surface and that the CBD solution has too fast homogeneous nucleation. In this deposition mode, the CdS has weak bonding with the CZTSSe surface, which can explain the lower bending performance of the control flexible cell. Comparatively, in the Rb sample, the CdS film is much more compact and smoother and exhibits much better morphology retention ability. This means that the heteroepitaxial deposition has been significantly improved. We consider that the CdS in the Rb sample realizes an ion-by-ion deposition mode (reaction function is shown in Figure S6) through sufficient ion adsorption and layer-by-layer reaction deposition. The more surface Zn vacancies provide sufficient atomic adsorption sites for the Cd ion and the broken Se-Se bond also facilitate the Cd adsorption through Cd-Se interactions. In this mode, more tightly bonded interface and more ordered CdS film microstructure can be obtained, which explains the better bending performance of the flexible cell.

Similar to the simple aqueous solution treatment, also no Rb in the film can be detected by XPS after the CdS deposition process, as in Figure 4(c). This further confirms that Rb regulation of the CZTSSe surface and the CdS deposition occurs in the solution and at the solution/solid interface while not realized through Rb insertion or doping. Raman spectra have been used to characterize the CZTSSe/CdS films. As in Figure 4(d-e), the Rb sample has a stronger and narrower CdS Raman signal at about 300 $cm^{-1}$.[47] In addition, the Raman peak intensity of the CZTSSe phase at about 196 $cm^{-1}$ is almost unchanged when increasing the CBD duration from 7 to 11 min, which is obviously different that of the control sample.[48]

This means that the coverage of the CZTSSe surface can be completed faster in the Rb sample due to the improvement in the CdS deposition mode. Further PL characterizations (Figure S7) also prove the quality of CZTSSe/CdS heterojunction has been much improved for much higher intensity. These results agree with the SEM result that the Rb regulation has significantly improved the compactness, smoothness and coverage of the deposited CdS film.

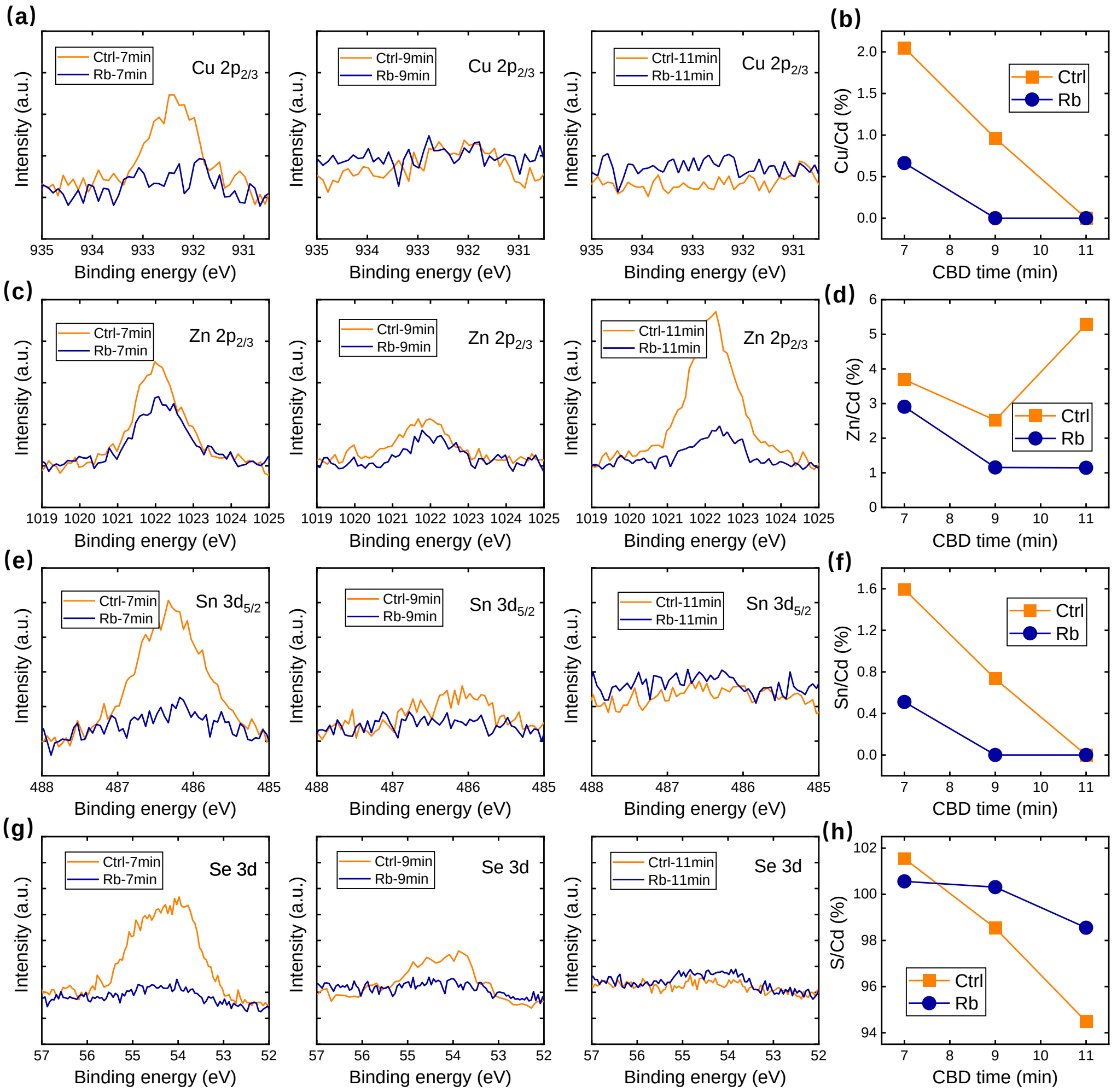

Figure 5. XPS spectra (Cu 2p, Zn 2p, Sn 3d and Se 3d) of the CZTSSe/CdS films at different CBD time points and the derived element content evolution.

The influence of Rb regulation on the CdS deposition can also be observed from the XPS measurement. As in Figure 5, the XPS intensity of Cu, Zn, Sn and Se elements is gradually decreased when increasing the CBD durations, which also reflects the coverage process of CdS. For the Rb sample, the ratio of Cu and Sn to Cd is rapidly decreased to 0 when the CBD time reaches 9 min, confirming a complete coverage of CdS on the CZTSSe film. Zn element can always be observed during the entire process and keeps a relatively constant atomic ratio to Cd (1%) after 9 min. This implies that Zn has incorporated into the CdS film through coprecipitation, which could form Zn/Cd substitutions. For the control sample, the ratio of Cu and Sn to Cd decreases much more slowly, as in Figure 5(h)&S8. In addition, Zn content shows an obvious fluctuation in the CdS deposition process, implying that the dissolving and precipitation behavior is not as stable as that in the Rb sample. The atomic ratio between S and Cd also indicates that the Rb sample has a more stable deposition behavior, while the control sample exhibits S deficiency in longer deposition time.[49] The stable deposition process is also a reason for the better CdS film quality.

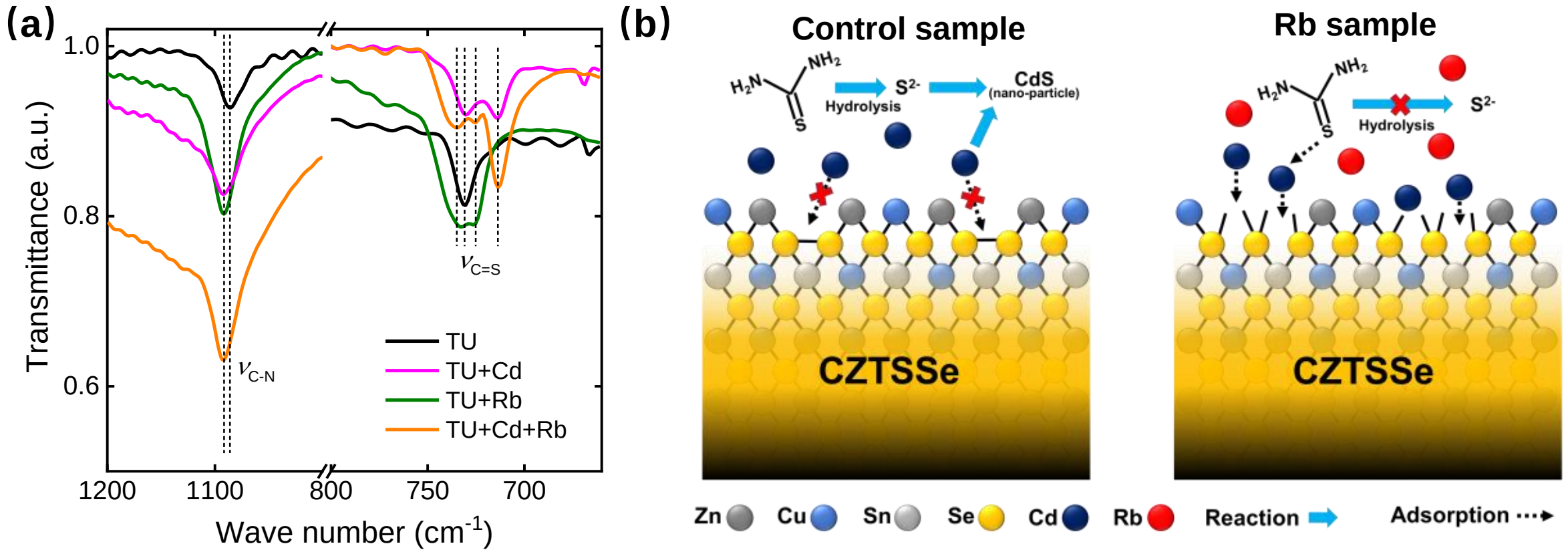

Figure 6. (a) FTIR spectra of the thiourea (TU), TU/Cd and TU/Rb mixtures derived from aqueous solutions. (b) schematic diagrams of the CdS CBD process and the influence of Rb ions.

To better understand the above phenomena, we further study the interactions between Rb ions and thiourea (TU). As in Figure 6(a), stretching vibrations of C-N ($\nu_{C\text{-}N}$) and C=S ($\nu_{C=S}$) in TU can be clearly observed in Fourier Transform infrared spectra (FTIR).[50] When Rb or Cd is combined with TU, the $\nu_{C\text{-}N}$ peak is a little shifted from 1086 to 1092 $cm^{-1}$, implying enhanced vibration of the C-N bond. For the $\nu_{C=S}$, it locates at about 731 $cm^{-1}$ in the pristine TU. After Cd is combined with TU, a new vibration peak appears at about 713 $cm^{-1}$ and the original peak still exists with a much weaker intensity.[51] This new lower-frequency vibration should be a signature of the coordination interaction between TU and Cd. Interestingly, when Rb is combined with TU, the original $\nu_{C=S}$ peak of TU disappears by splitting into two new peaks, locating at 734 and 725 $cm^{-1}$, respectively. This means that TU has sufficiently coordinated with Rb ions. When Cd and Rb are combined together with TU, these three new vibration peaks all appear and additionally the 713 $cm^{-1}$ peak intensity become obviously stronger compared to the TU-Cd sample. This result implies that the existence of Rb ions can help eliminate free TU by coordination and can also promote the Cd-TU coordination. Ion exchange in the TU-M (M: Cd or Rb) complexes may be the mechanism for this phenomenon.[52]

Combining with the schematic diagram of the CdS deposition process and above experimental results, here we will give a discussion of the possible mechanism for the improvement in the CdS deposition and heterojunction interface quality. As in Figure 6(b), Se atom in the control CZTSSe lattice surface prefers to bond to another Se, forming $Se^0$ deep surface defect. $Se^0$ structure also has low activity to coordinate with Cd ions in the CBD solution, thus affecting the ion adsorption and CdS heteroepitaxial deposition. In addition, free TU in the alkaline CBD solution can hydrolyze into $S^{2-}$, which will coprecipitate with $Cd^{2+}$ to form CdS particles and thus induce the cluster-by-cluster CdS deposition mode. Apparently, the $Se^0$ surface defect and the cluster-by-cluster deposited CdS layer is not beneficial for obtaining high-quality heterojunction interface. Comparatively, in the Rb sample, alkali metal-Se interaction helps to break up the covalent bond in $Se^0$ structures and thus make the surface Se atom be active to interact with Cd ions. The formation of Cd-Se bond on the CZTSSe surface on one hand can facilitate the CdS heteroepitaxial deposition, obtaining high interface robustness, and on the other hand can eliminate the surface defect state. In addition, the coordination interaction between Rb and TU can also stabilize the molecular structure of TU in the alkaline CBD solution, which thus helps suppress the homogeneous nucleation of CdS and avoids the cluster-by-cluster film growth.

**Conclusion**

In this work, we have introduced alkali metal ions in the CdS deposition process to improve the heterojunction interface quality of flexible CZTSSe solar cells. In the Rb ion regulated cell, high total-area PCE of 12.63% (active-area efficiency) and excellent bending performance has been achieved. The performance parameter of the cell has been amongst the highest results reported so far and is even comparable to that of conventional rigid solar cells. Photoelectric property characterization shows that this high performance mainly comes from better interface charge transport and lower interface charge recombination. Material and chemical characterization has found that the Rb ion regulation can break up the Se-Se bond on the CZTSSe lattice surface, which on one hand can passivate the $Se^0$ surface deep defect and one the other hand can promote the Cd-Se interaction and CdS heteroepitaxial deposition. The Rb ion is also found to be able to coordinate with TU in the CBD solution, stabilize the molecular structure of TU and thus promote the ion-by-ion deposition of CdS layer. The synergistic effect of these benefits finally results in efficient photoelectric conversion and robust cell interfaces. Overall, this work realizes impressive progress in flexible CZTSSe solar cells and provides valuable insights into surface and interface regulation of CZTSSe solar cells.

## Methods

*Reagents and Materials*: thiourea (99%, Alfa), 2-Methoxyethanol (EGME) (99.8%, Aladdin), AgCl (99.5%, Innochem), CuCl (99.99%, Alfa), $SnCl_4$ (99.998%, Macklin), $Zn(Ac)_2$ (99.99%, Aladdin), Molybdenum foil (≥99.95%, 0.1 mm thick, Qinghe County Haoxuan Metal Materials Co., Ltd.)

*CZTSSe precursor film preparation*: Firstly, 7.311 g thiourea was added into Vial 1 containing 15 ml EGME, and stirred until dissolved. Then, 0.345 g AgCl, 2.16 g CuCl were successively added into Vial 1, stirred till completely dissolved. Secondly, 15 ml EGME was injected into the Vial 2 containing 3.963 g $SnCl_4$ under stirring. Thirdly, 3.126 g $Zn(Ac)_2$ was added into the $SnCl_4$-EGME suspension till thoroughly dissolved. Fourthly, mixed the solution in Vial 2 and Vial 1, then obtained clear precursor solution. All above were performed in glove box.

The filtered precursor solution was spin-coated onto a pre-cleaned Mo substrate at 2000 rpm for 30s, followed by annealing on a hot plate at 280°C for 2 min. This coating-annealing process was repeated 4 times. Then, precursor films were placed in a graphite box containing Se particles and selenized in a rapid heating tube furnace. The detailed selenization condition was as followed: the temperature was raised to 535 °C within 1 min and maintained for 20 min. The whole selenization process was performed under one atmosphere with $N_2$ flow of 80 sccm.

*CZTSSe device fabrication*: A 40~50 nm thickness CdS buffer layer was deposited on the top of CZTSSe films by the chemical bath deposition (CBD) method, followed by sputtering 50 nm i-ZnO layer and 200 nm ITO layer. 50 nm nickel (Ni) and 2 μm aluminum (Al) were evaporated on ITO layer. 110nm $MgF_2$ layer covered the whole device, serving as the anti-reflection coating (ARC). Finally, the total area of each cell is 0.28 $cm^2$, which was measured by optical microscope.

*Film Characterization*: Raman spectra were carried out on Raman spectrometer (Lab-RAM HR Evolution, HORIBA) by using 532 nm laser diode as the excitation source. FTIR characterizations were performed by a Fourier Transform Infrared (FTIR) Spectrophotometer (TENSOR27, Bruker). Scanning electron microscopy (SEM) images were measured on Hitachi S4800 SEM using 10 kV power. Kelvin probe force microscope (KPFM) images were obtained on an atomic force microscope (Multimode 9, Bruker). The XPS characterizations were performed on X-ray photoelectron spectrometer (Thermo Fisher Scientific ESCALAB 250Xi). Optical images of samples were measured on metallographic microscope (OLYMPUS, BX61). Steady-state photoluminescence (PL) spectra were obtained from PL spectrometer (Edinburgh Instruments, FLS 900), excited with a picosecond pulsed diode laser (EPL-640) with the wavelength of 638.2 nm while cooling down with liquid helium.

*Device Characterization*: Modulated transient photocurrent and photovoltage (M-TPC/TPV) measurements were obtained by our lab-made setup, in which the cell was excited by a tunable nanosecond laser pumped at 532 nm and recorded by a sub-nanosecond resolved digital oscilloscope (Tektronix, DPO 7104) with a sampling resistance of 50 Ω or 1 MΩ. The current density-voltage (*J*-*V*) curves were recorded on Keithley 2400 Source Meter under simulated AM 1.5 sunlight at 100 mW $cm^{-2}$ calibrated with a Si reference cell (calibrated by NIM). *J*-*V* test was conducted under 25 °C in air. Its scanning speed is 130 mV/s and there is 0.1s dwell time at each sampling voltage. Each statistical box of device performance in this paper contains data of 9 cells. External quantum efficiency (EQE) was measured by Enlitech QE-R test system using calibrated Si and Ge diodes as references. The drive-level capacitance profiling (DLCP) was measured on an electrochemical workstation (Versa STAT3, Princeton) by using 11 kHz and 100 kHz AC excitation with amplitude from 10 to 100 mV and with DC bias from 0 to -0.4 V. Temperature dependent *J-V* data was collected by electrochemical workstation (Versa STAT3, Princeton) under LED light source (S3000, Nanjing Hecho Technology Co., Ltd), and temperature was controlled by temperature controller (Model 336, Lake Shore Cryotronics, Inc.).

*Bending tests*: The bending tests are realized via bending flexible device to fit the surface of wooden cylinder. And the bending radius is determined by the size of cylinder. In bending radius test, all devices maintain bending state for 10 s. In durability tests, the maintaining time is 1s within each bending cycle.